\documentclass[aps,pra,twocolumn,showpacs,floatfix,superscriptaddress]{revtex4-1}

\usepackage{mathptmx} 
\usepackage{dsfont} 
\usepackage{bm} 

\usepackage{times}
\usepackage{physics}
\usepackage{graphicx}
\usepackage{hyperref}
\hypersetup{colorlinks=true,linkcolor=blue,citecolor=blue}
\usepackage{amsmath,amssymb,amsfonts}
\usepackage{wrapfig}
\usepackage{siunitx}

\def\ket#1{\mathinner{|{#1}\rangle}} 

\begin{document}

\title{Hybrid systems for the generation of non-classical mechanical states via quadratic interactions}

\begin{abstract}
We present a method to implement two-phonon interactions between  mechanical resonators and  spin qubits in hybrid setups, and show that these systems can be applied for the generation of nonclassical mechanical states even in the presence of dissipation. In particular, we demonstrate that the implementation of a two-phonon Jaynes-Cummings Hamiltonian under coherent driving of the qubit yields a dissipative phase transition with similarities to the one predicted in the model of the degenerate parametric oscillator: beyond a certain threshold in the driving amplitude, the driven-dissipative system sustains a mixed steady state consisting of a ``jumping cat'', i.e., a cat state undergoing random jumps between two phases.  We consider realistic setups and show that, in samples within reach of current technology, the system features non-classical transient states, characterized by a negative Wigner function, that persist during timescales of fractions of a second.
\end{abstract}
\date{\today}

\author{Carlos \surname{S\'anchez Mu\~noz}}
\altaffiliation{Present address: Clarendon Laboratory, University of Oxford, Oxford OX1 3PU, UK}
\email{carlossmwolff@gmail.com}
\affiliation{Theoretical Quantum Physics Laboratory, Cluster for Pioneering Research, RIKEN, Saitama, 351-0198, Japan}
\author{Antonio Lara}
\affiliation{Dpto. F\'isica Materia Condensada C03, Instituto Nicolas Cabrera (INC), Condensed Matter Physics Institute (IFIMAC), Universidad Aut\'onoma de Madrid, 28049, Madrid, Spain}
\author{Jorge Puebla}
\affiliation{Theoretical Quantum Physics Laboratory, Cluster for Pioneering Research, RIKEN, Saitama, 351-0198, Japan}
\author{Franco Nori}
\affiliation{Theoretical Quantum Physics Laboratory, Cluster for Pioneering Research, RIKEN, Saitama, 351-0198, Japan}
\affiliation{Department of Physics, University of Michigan, Ann Arbor, MI 48109-1040, USA}
\maketitle

\emph{Introduction.}---Over the past decades, technological developments have allowed to implement new classes of extremely sensitive nanomechanical oscillators,  such as membranes or microcantilevers, that are finding applications in a  wide variety of areas, from biological detection~\cite{calleja12a} to ultrasensitive mass sensing~\cite{spletzer08a,gil09a,chaste12a} or NMR imaging~\cite{mamin07a,poggio10a,mamin12a}. There has been a growing interest in studying hybrid systems in which these mechanical elements are coupled to some other quantum actor, allowing to explore the quantum limits of mechanical motion~\cite{wallquist09a,xiang13a,treutlein_book14a}, with prominent examples such as cavity optomechanics setups~\cite{gigan06a,thompson08a,kippenberg08b,marquardt09a,aspelmeyer14a}. Many of these works aim to explore the quantum limit of mesoscopic objects consisting of billions of atoms by cooling them close to the ground state~\cite{blencowe04,chan11a,poot12a} and generating inherently quantum states, such as squeezed states~\cite{jahne09} or quantum superpositions~\cite{hoff16a}. 

In this work, we present hybrid setups that are able to achieve a two-phonon coherent coupling between a mechanical mode and a spin qubit, described as a two-level system (TLS).
It is known for systems involving some kind of two-particle interaction plus a nonlinearity~\cite{
drummond80b,drummond81a,gilles94a,hach94a,benito16a,krippner94a,hu96a,hu97a,hu99a,nation12a,tan13a,everitt14a,mirrahimi14a,leghtas15a,minganti16a,bartolo16a},
that the mechanical system can evolve into motional cat states. Although these states are ultimately washed out by decoherence, our proposed setup features non-classical transient states, characterized by a negative Wigner function, during timescales that can extend up to seconds. After this, the system reaches a mixed steady state that has been understood~ as a cat state flipping its phase at random times~\cite{minganti16a}. This offers an attractive platform both for the study of fundamental questions in quantum mechanics---such as decoherence, spontaneous symmetry breaking and ergodicity in dissipative quantum systems~\cite{zurek03a,molmer97a,cresser2001ergodicity,benito16a}---and for practical applications where nonclassical mechanical states can be envisaged as a technological resource~\cite{ralph03,lund08,joo11,facon16a,albert16a,mirrahimi14a} .

Our proposal is based on hybrid devices in which a single spin qubit embedded in a magnetic field gradient couples to a mechanical oscillator through the position-dependent Zeeman shift~\cite{rabl09a,rabl10a,arcizet11a,kolkowitz12a,pigeau15a,wei15a,li16a}. We consider the qubit to be given by the electronic spin of nitrogen-vacancy (NV) centers, which are excellent candidates due to their outstanding coherence and control properties~\cite{childress06,gaebel06a,hanson06a,hanson08a,buluta11a,doherty13a}. Several works~\cite{ma16a,cai17a} have analyzed particular geometries in which the equilibrium position of the system leaves the spin in a point of null magnetic gradient, leading to a quadratic dependence of the coupling with the position. These works proposed to use this dependence to couple two different modes of the oscillator in order to effectively enhance  the linear coupling between one of these modes and the TLS. In contrast, we propose here to use these geometries to achieve degenerate, two-phonon exchange between one mode of the resonator and the TLS, which gives rise to physical phenomena with no analogue in linearly-coupled systems~\cite{wang17a}.

\begin{figure}[t!]
\begin{center}
\includegraphics[width=1\columnwidth]{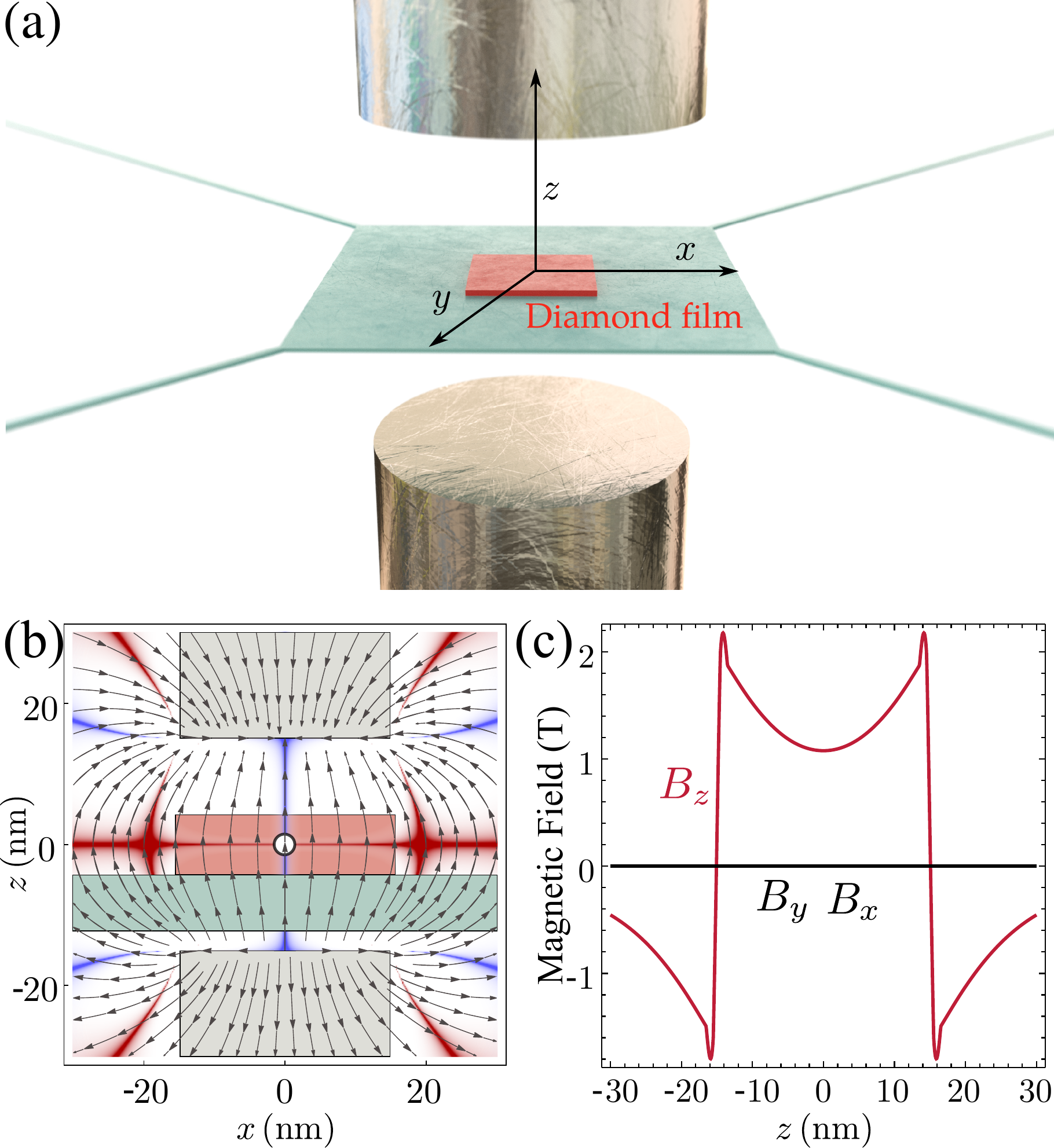}
\end{center}
\caption{(a) Proposed configuration: an NV center in a diamond film (red) is placed on top of a mechanical oscillator (e.g., a thin membrane) and placed  between two  magnets with aligned magnetization. Dimensions have been altered for visual clarity. (b) Magnetic field lines, computed for two Dy magnetized cylinders of 30~nm diameter separated by a gap of \num{30}~nm. Red (blue)-shaded areas mark the regions where $\partial B_{z(x)}/\partial z\approx 0$. Both first derivatives are zero at the position of the NV center---represented by a circle---when the oscillator is at rest. (c) Magnetic field along the $z$ axis for $x=0$.}
\label{fig:1}
\end{figure}
\emph{Setup proposal.}---We consider an NV center placed on top of a mechanical oscillator at a position $\mathbf{r}_0$ and surrounded by a magnetic field $\mathbf{B}(\mathbf{r})$. An NV center consists of a nitrogen atom and an adjacent vacancy in diamond, and its electronic ground state can be described as a $S=1$ spin triplet with states $|m_s\rangle$, with $m_s = 0, \pm 1$. The Hamiltonian of the system reads (we set $\hbar = 1$ hereafter) $H =H_\mathrm{NV}+H_\mathrm{M} +  \mu_\mathrm{B} g_s \,\mathbf{S}\cdot \mathbf{B}(\mathbf{r}_0)$, where
$H_\mathrm{NV}$ stands for the Hamiltonian of the NV center, $H_\mathrm M$ for the mechanical mode, and the last term describes a perturbation on the NV center due to the external magnetic field, where $\mu_\mathrm{B}$ is the Bohr magneton, $g_s = 2$ is the Land\'e factor of the NV center, and $\mathbf{S}$ is its spin operator. The last term provides the mechanism that couples the qubit and the mechanical degree of freedom. We will assume that the mechanical mode oscillates only along the $z$ axis, so that the position of the NV center is given by $\mathbf{r}_0=(0,0,z)$, $z$ being the displacement of the oscillator with respect to the equilibrium point. Setting  $\mathbf{B}(\mathbf{r}_0) \equiv \mathbf{B}(z)$, we can expand the Hamiltonian in terms of $z$ up to second order, $H \approx H_\mathrm{NV}+H_\mathrm{M} + \mu_\mathrm{B} g_s \mathbf{S} \cdot \left[ \partial \mathbf{B}/\partial z(0)(0)z +\frac{1}{2}\partial^2 \mathbf{B}/\partial z^2(0){z}^2 \right]$. 
Our proposal relies on considering a magnetic field with an extremum at the position of the NV center, which will cancel the first derivative in the expansion and provide a second-order coupling to the mechanical mode. 
For simplicity, we will consider that the field has also null second derivatives along the $x$ and $y$ axis, so that the mechanical mode only couples to $S_z$. This latter assumption is not necessary, but we show here a particular proposal in which this is indeed the case. By writing the position operator of the mechanical oscillator as $z = z_\mathrm{zpf} (a+a^\dagger)$, with ${z_\mathrm{zpf} = \sqrt{\hbar/(2 m_\mathrm{eff}\omega_m)}}$ the zero-point fluctuation amplitude, $\omega_m$ the resonant mechanical frequency, and $a$ the annihilation operator, the resulting Hamiltonian becomes:
\begin{equation}
H \approx H_\mathrm{NV}+H_\mathrm{M}  + g_2 \, (a^\dagger + a)^2  S_{z}\, ,
\end{equation}
where the two-phonon coupling is given by ${g_2 = \frac{1}{2}\mu_\mathrm{B} g_s \, {z_\mathrm{zpf}}^2 G_2}$, and ${G_2 = \partial^2 B_z/\partial z^2(0)}$. The critical parameters here in order to maximize this coupling are the second gradient of the magnetic field and the zero-point motion of the oscillator, which in both cases should be as high as possible.

For practical purposes, we will focus on cases where the magnetic field is generated by nanomagnets, since they provide high gradients at short distances~\cite{mamin07a,poggio10a,mamin12a}.
In Fig.~\ref{fig:1}, we propose a particular arrangement of magnets that yields the required spatial magnetic field profile. An NV center injected in a diamond film~\cite{ohashi13a} is placed on top of a resonator of nanometer-scale thickness that oscillates along the $z$ direction; 
the extension of the diamond film should be much smaller than that of the oscillator to minimize any possible impact on its properties. Diamond films can be compatible, for instance, with silicon nitride substrates~\cite{almeida11a,baudrillart16a,skoog12a}.  The resonator is positioned in the gap between two cylindrical nanomagnets with saturated magnetization  along the $z$ axis. The size of the gap is considered to be of the order of tens of nanometers. In the region between the magnets, this geometry yields a strong magnetic field in the $z$ direction and a negligible field in the $x$ and $y$ directions, as we show in Fig~\ref{fig:1}(b-c). Moreover, every component of the field has a null derivative with respect to $z$ at the middle point. This gives rise to the quadratic coupling between the NV center and the oscillator. 
 
\begin{figure}[t!]
\begin{center}
\includegraphics[width=0.9\columnwidth]{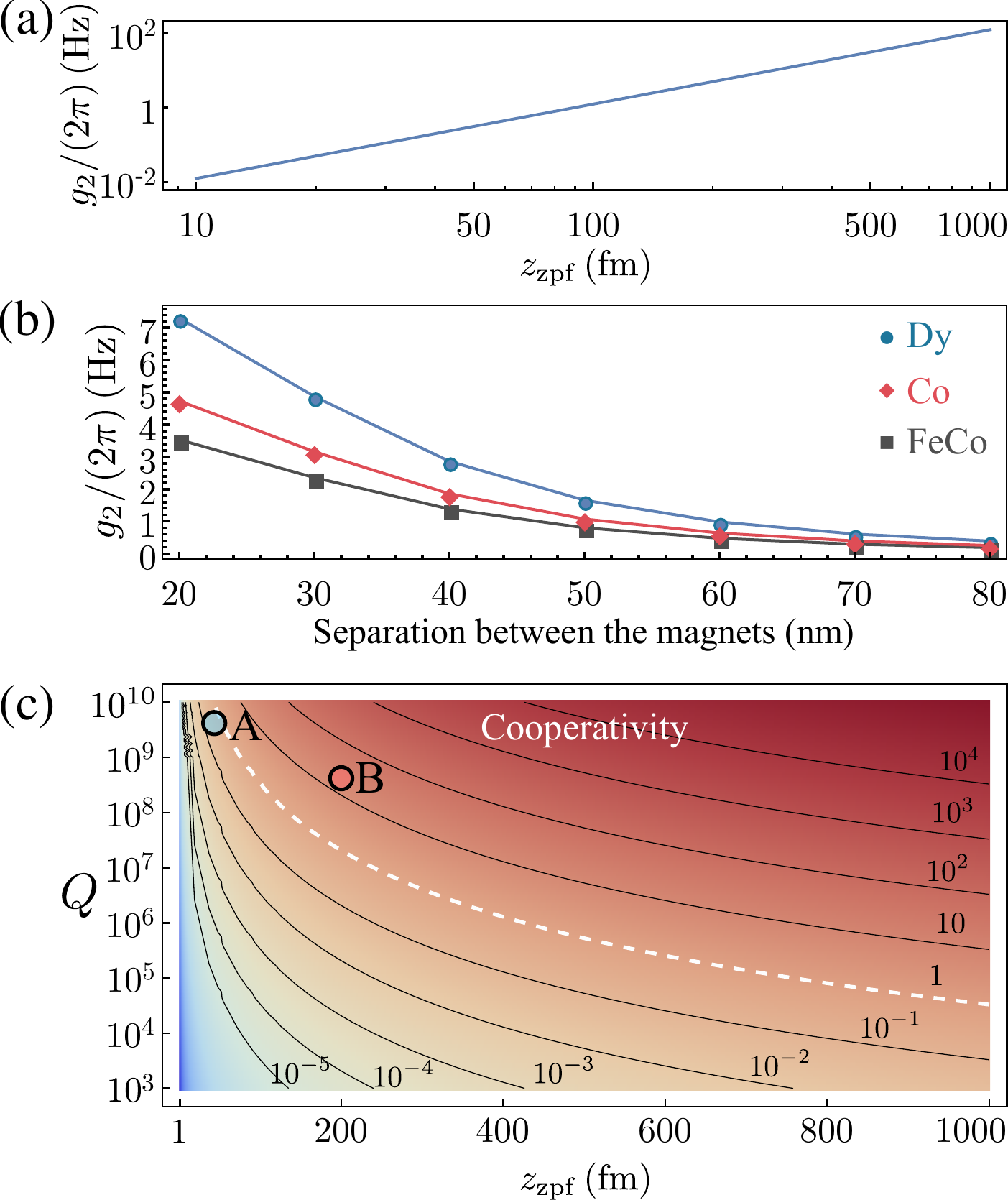}
\end{center}
\caption{(a) Two-photon coupling rate versus zero-point fluctuation amplitude $z_\mathrm{zpf}$ of the oscillator, for two $\mathrm{Dy}$ magnets separated by \num{30}~nm. (b) Two-photon coupling rate versus the magnet separation for three different magnetic materials, for a resonator with $z_\mathrm{zpf}=$~\num{200}~fm. (c) Cooperativity versus the oscillator quality factor $Q$ and the zero-point fluctuations, for a magnet separation of \num{30}~nm, oscillator frequency $\omega_m/(2\pi) = $~\SI{1.8}{\mega\hertz}, temperature $T=$~\SI{10}{\milli\kelvin}, and pure-dephasing rate $\gamma_z/(2\pi)=$~\SI{10}{\hertz}. The white line $C=1$ marks the onset of quantum effects. Point A corresponds to a feasible point for state-of-the-art technology at mK temperatures~\cite{ghadimi18a}, with $z_\mathrm{zpf}=$~\num{43} fm and $Q=$~\num{4.2e9}; B is the point taken in most part of the text for clarity of results: $z_\mathrm{zpf}=$~\num{200} fm and $Q=$~\num{4.2e8}}
\label{fig:2}
\end{figure}

\emph{Two-phonon coupling rates.}---In order to estimate the achievable two-photon coupling rate in realistic setups, we simulated the magnetic field generated by two cylinders of nanometer size with saturated magnetization for three different materials ($\mathrm{Dy}$, $\mathrm{Co}$ and $\mathrm{FeCo}$) (see Appendix~\ref{sec:apA}).  $\mathrm{Dy}$ stands as the best choice due to its high saturation magnetization~\cite{mamin12a,scheunert14a}. Figure~\ref{fig:2}(b-c) is an example of the simulated magnetic field for two cylinders of $\mathrm{Dy}$ with \num{30}~nm  of diameter, \num{150}~nm of height and separated by a gap of \num{30}~nm. In this configuration, one can obtain values of $G_2\approx$~\SI{9e15}{\tesla\meter^{-2}}.  To obtain the corresponding two-phonon coupling rate $g_2$, one must consider a specific implementation of the mechanical oscillator. The relevant parameter is the zero-point fluctuation amplitude, $z_\mathrm{zpf}$, which ranges from tens of femtometers in systems such as $\mathrm{Si}_3\mathrm{N}_4$ oscillators~\cite{norte16a,ghadimi18a} to hundreds of femtometers in  systems such as carbon nanotubes~\cite{huttel09a}, graphene resonators~\cite{singh14a,weber16a}, $\mathrm{SiC}$ wires~\cite{arcizet11a} or $\mathrm{Si}$ cantilevers~\cite{poggio07a}. Figure~\ref{fig:2}(a) shows $g_2$ versus $z_\mathrm{zpf}$ for a fixed separation between $\mathrm{Dy}$ magnets of \num{30}~nm. As an example, for a value of $z_\mathrm{zpf}\approx$~\num{200}~fm, the resulting coupling rate is $g_2/(2\pi)\approx$~\SI{5}{\hertz}. The dependence of this value on the gap between the magnets is shown on Fig.~\ref{fig:2}(b). Linear couplings induced by non-zero first-order gradients due to imperfect alignment can be disregarded at the two-phonon resonance condition. Alternatively, they could be used to calibrate the device and measure the state of the oscillator by properly tuning the qubit energy (see Appendix~\ref{sec:apD}-\ref{sec:apE}).
 
\emph{Quantum effects.}---To address the  possibility of observing quantum effects, a relevant figure of merit is the cooperativity $C = 4g_2^2/[\gamma_z \gamma_m(n_\mathrm{th}+1) ]$~\cite{kolkowitz12a,schuetz15a}, where $\gamma_z$ is the dephasing rate of the qubit, $\gamma_m=\omega_m/Q$ is the oscillator decay rate ($Q$ being the quality factor), and $n_\mathrm{th}$ is the average number of thermal phonons at the oscillator at the temperature $T$. Values of the cooperativity $C >1$ mark the onset of quantum effects. 
The impact of spin relaxation is not relevant here, since relaxation times can reach hundreds of seconds at low temperatures~\cite{jarmola12}. Regarding pure dephasing rates, $\gamma_z$ can achieve room temperature values $\sim$\SI{1}{\hertz}~\cite{bar13a} using dynamical decoupling techniques, which have already been used in very similar setups~\cite{kolkowitz12a}.
Once $\gamma_z$, $T$ and $G_2$ are established, the cooperativity $C$ is fully determined
by the oscillator parameters, $\omega_m$, $Q$ and $z_\mathrm{zpf}$. Figure~\ref{fig:2}(c) shows $C$ versus $Q$ and $z_\mathrm{zpf}$ for $\omega_m\sim$\SI{}{\mega\hertz} (typical of systems such as $\mathrm{SiC}$ wires~\cite{arcizet11a} or $\mathrm{Si}_3\mathrm{N}_4$ nanobeams~\cite{ghadimi18a}), $\gamma_z/(2\pi)=$~\SI{10}{\hertz}  and $T=$~\SI{10}{\milli\kelvin}.  
As an example, an oscillator with $\omega_m/(2\pi)=$~\SI{1.8}{\mega\hertz},  ${z_\mathrm{zpf}\approx 43}$~fm~\cite{ghadimi18a} and $Q\approx$~\num{4e9} (point A in Fig.~\ref{fig:2}(c)) yields $C\approx 0.4$ at these conditions, and can show quantum effects for dephasing rates $\gamma_z /(2\pi)<$~\SI{4.3}{\hertz}, which have already been achieved experimentally~\cite{bar13a,ovartchaiyapong14a}. 
Recently, room-temperature values $Q>10^8$ have been demonstrated in oscillators fabricated via soft-clamping and strain engineering techniques~\cite{tsaturyan17a,ghadimi18a}, with values $Q>10^9$ expected at dilution refrigerator temperatures (\SI{14}{\milli\kelvin})~\cite{tsaturyan17a}. Therefore, although demanding, these conditions are withing reach of state-of-the-art technology. 
For the sake of clarity of results, we will consider hereafter a slightly more optimistic value of $z_\mathrm{zpf}\approx$~\num{200}~fm (giving $g_2/(2\pi)=$~\SI{5}{\hertz}), and set $Q =$~\num{4.2e8} and $\omega_m/(2\pi) =$~\SI{1.8}{\mega\hertz} as in Ref.~\cite{ghadimi18a} (this choice is shown as point B in Fig.~\ref{fig:2}(c)). We take $\gamma_z/(2\pi) =$~\SI{10}{\hertz} and $T = $\SI{10}{\milli\kelvin}, giving $n_\mathrm{th} \approx 115$ and $C \approx 20$. While the proximity  of the NV center to the surface in a diamond film might render longer dephasing rates than in the bulk, we note that we are also considering cryogenic temperatures, which is known to enhance coherence times by several orders of magnitude~\cite{jarmola12}. At these low temperatures, several techniques exist in order to minimize the influence of heat induced by, e.g., RF voltage; most of these solutions are related to the design of heat sinks, cooling fins, etc., and the selection of proper materials for heat dissipation~\cite{savin06a}.

\begin{figure}[t!]
\begin{center}
\includegraphics[width=0.9\columnwidth]{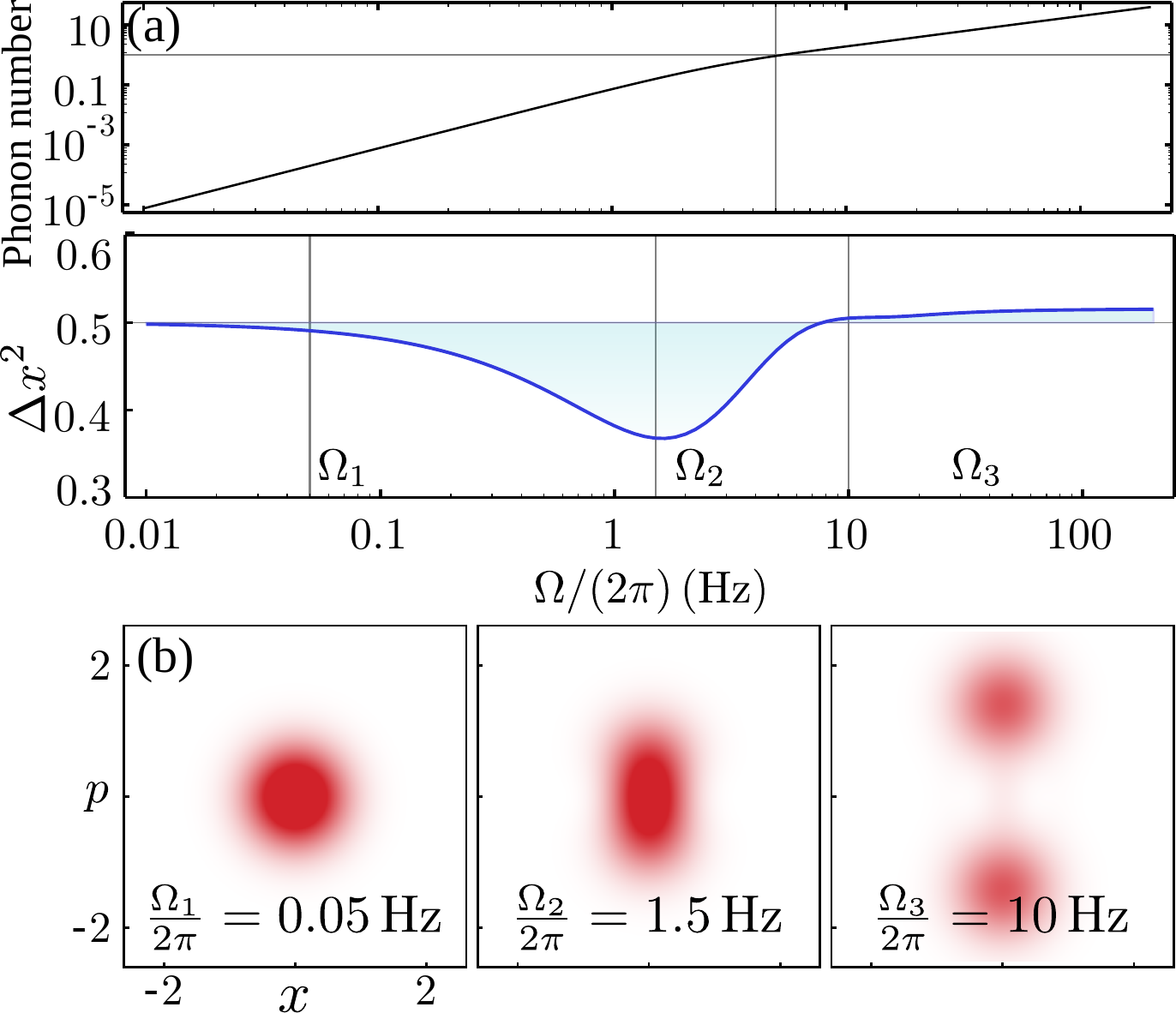}
\end{center}
\caption{Signatures of a dissipative phase transition in the steady-state. (a) Phonon number and fluctuations of the position operator versus the driving amplitude $\Omega$. (b) Wigner function for three values of the driving amplitude.    Here, {$g_2/(2\pi)=$~\SI{5}{\hertz}}, $Q=$~\num{4.2e8}, $\omega_m/(2\pi)=$~\SI{1.8}{\mega\hertz}, $T=$~\SI{10}{\milli\kelvin}, $\gamma_z/(2\pi)=$~\SI{10}{\hertz}.}
\label{fig:3}
\end{figure}
\begin{figure}[t!]
\begin{center}
\includegraphics[width=0.95\columnwidth]{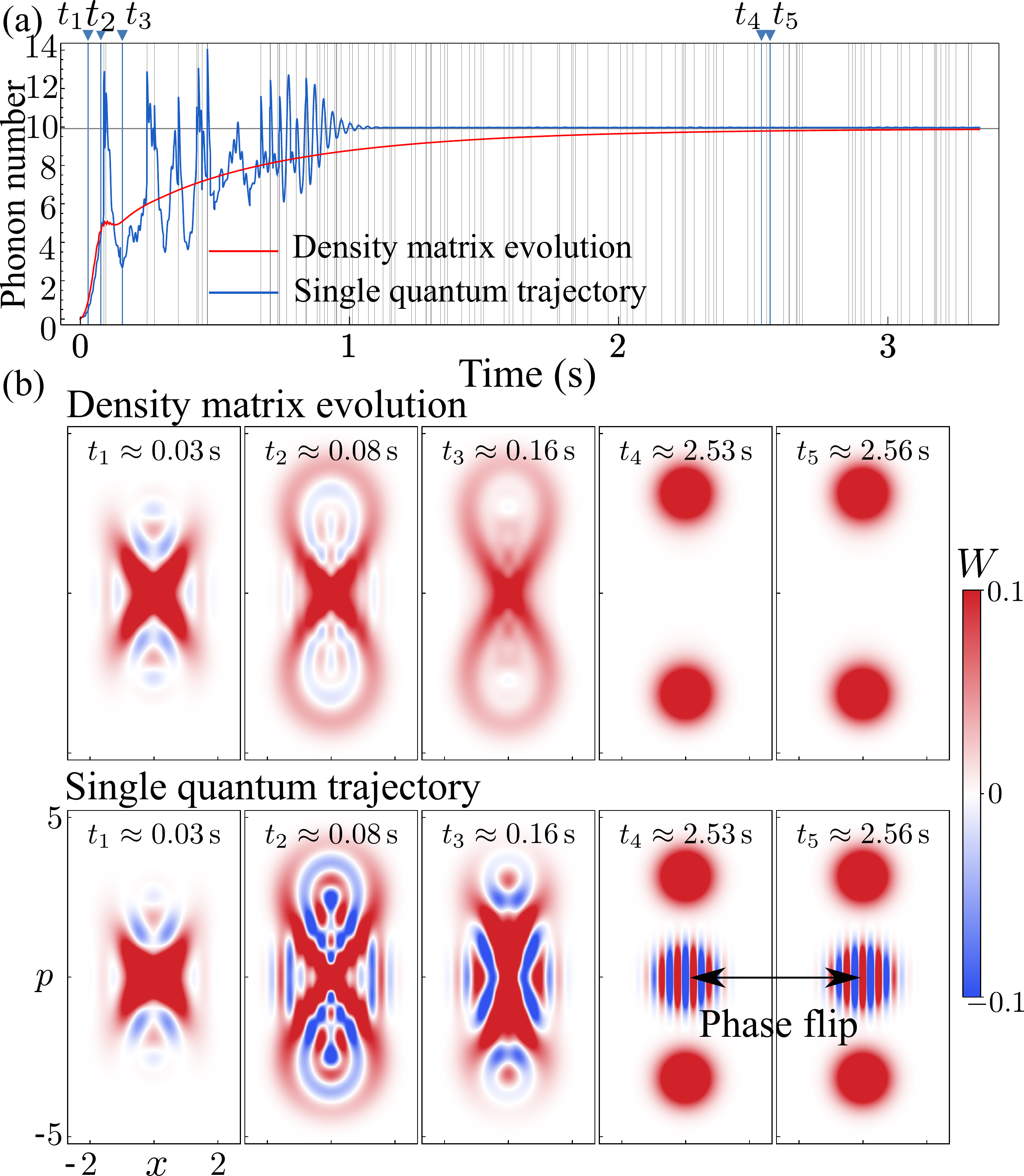}
\end{center}
\caption{Dynamics of the density matrix and of a single quantum trajectory. (a) Phonon population versus time. Vertical lines depict times when a phonon emission process takes place. (b) Wigner function of the mechanical mode given by the density matrix (upper) and the wavefunction of a single trajectory (lower), at the times marked with arrows on the top of panel (a). The last two columns of the bottom row depict a ``jumping cat'', at times before and after a phonon emission process. Parameters are those of Fig.~\ref{fig:3}.}
\label{fig:4}
\end{figure}
\emph{Dissipative dynamics of the driven, two-phonon Jaynes-Cummings Hamiltonian.}---By adding two oscillating magnetic fields, one in the $x$ axis with frequency $\omega_x$ in the MW regime; and another in the $z$ axis with frequency $\omega_z \sim \omega_m$, we obtain (see Appendix~\ref{sec:apC} and Refs.~\cite{rabl09a,li16a}) an effective, coherently driven two-phonon Jaynes-Cummings Hamiltonian:
\begin{multline}
H = (\omega_\sigma-\omega_z)\, \sigma^\dagger\sigma + (\omega_m-\omega_z/2) a^\dagger a +
\\ \Omega(\sigma+\sigma^\dagger ) + g_2({a^\dagger}^2\sigma + a^2 \sigma^\dagger),
\label{eq:2-phonon-JC}
\end{multline}
where $\sigma$ is the lowering operator of the effective TLS, and $\Omega$ denotes the amplitude of the driving. We will consider the resonant situation $\omega_\sigma = 2\omega_m$.
In order to describe the dynamics of the system under dissipation, this Hamiltonian needs to be supplemented with the usual Lindblad terms~\cite{carmichael_book02a}, giving the master equation for the dynamics of the density matrix, $\dot{\rho} = -i[H,\rho] + (\gamma_m n_\mathrm{th}/2)\mathcal{L}_a[\rho] + (\gamma_z/2)\mathcal{L}_{\sigma^\dagger\sigma}[\rho]$, 
where $\mathcal{L}_O[\rho] \equiv 2 O\rho O^\dagger - O^\dagger O\rho - \rho O^\dagger O$. We consider the system to be actively cooled to a thermal phonon population close to zero, which can be done, for instance, by means of laser cooling~\cite{chan11a,teufel11a,peterson16a} or using another spin qubit~\cite{rabl09a}. We therefore exclude incoherent pumping terms of the kind $\mathcal L_{a^\dagger}$ from the master equation, at the expense of using an increased resonator linewidth $\gamma_m n_\mathrm{th}$, with $\gamma_m$  the natural linewidth, and $n_\mathrm{th}$ is the number of thermal phonons in the oscillator in the absence of cooling~\cite{kolkowitz12a}.

The two-phonon Hamiltonian \eqref{eq:2-phonon-JC} is reminiscent of quantum optical systems with two-photon interactions  that have attracted considerable interest~\cite{gilles94a,hu96a,hu97a,hu99a,everitt14a,mirrahimi14a,leghtas15a}. Different systems with two-particle interactions and some kind of nonlinearity---e.g., two-photon losses in the case of the degenerate parametric oscillator (DPO)~\cite{drummond80b,drummond81a,gilles94a,hach94a,benito16a,nation12a}, a Kerr nonlinearity~\cite{goto16a} or, as in the present case, a TLS~\cite{wang17a}---,  have been shown to develop transient cat states~\cite{krippner94a,tan13a,everitt14a,leghtas15a,goto16a,wang17a} that, through unavoidable single-photon losses, tend to a steady state characterized by a Wigner function with phase bimodality~\cite{gilles94a,hach94a,benito16a,bartolo16a} and no interference fringes. Research on the DPO has shown that such steady state corresponds to a succession of random jumps between cat states of opposite phase when single trajectories are considered~\cite{minganti16a}, i.e., a sustained ``jumping cat''~\cite{garraway94}.
In the following, we discuss the appearance of analogue nonclassical effects in our system.

\begin{figure}[t!]

\begin{center}
\includegraphics[width=1.\columnwidth]{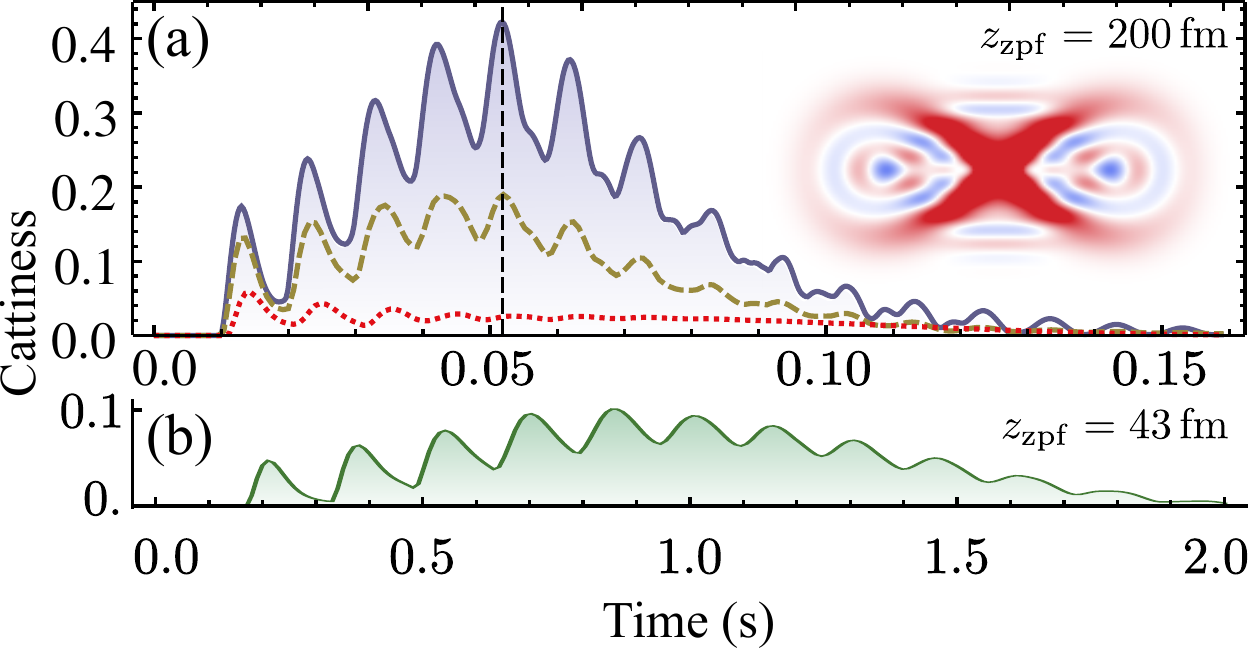}
\end{center}
\vspace{-10pt}
\caption{Cattiness $\mathcal C$ versus time for an initial vacuum state. (a)
Parameters are those of Fig.~\ref{fig:3}, with $\gamma_z/(2\pi) = $~\SI{10}{\hertz} (solid, blue), \SI{20}{\hertz} (dashed, yellow) and \SI{50}{\hertz} (dotted, red).
Inset: Wigner function at the time of maximum $\mathcal C$, $t\approx 51$~ms. (b) $\Omega/(2\pi)=$~\SI{3.18}{\hertz} and parameters of point A in Fig.~\ref{fig:2}(c), consistent with  Ref.~\cite{ghadimi18a} at $T=$~\SI{10}{\milli\kelvin}:  $z_\mathrm{zpf}=43$~fm ($g_2\approx$~\SI{0.23}{\hertz}), $Q=$~\num{4.2e9} (estimated), and $\gamma_z/(2\pi)=$~\SI{1}{\hertz}, giving $C\approx 4$. }
\label{fig:5}
\vspace{-10pt}
\end{figure}

Figure~\ref{fig:3}(a) depicts the phonon population and the variance of the position operator in the steady state versus the driving amplitude $\Omega$. In close similarity to the DPO~\cite{drummond80b,drummond81a,gilles94a,hach94a,benito16a}, we observe a phase transition characterized by the development of two lobes in the Wigner function, preceded by some degree of squeezing. This occurs when the phonon population is $\approx 1$, a point where its dependence with $\Omega$ changes from $\propto \Omega^2$ to $\propto \Omega$. Note that here, phase bimodality does not originate from the two-level nature of the driven TLS~\cite{alsing91a}, but is rather a consequence of the phase symmetry of the master equation, which is invariant under the change $a\rightarrow -a$~\cite{benito16a}.  Figure~\ref{fig:4} shows the transient dynamics of the oscillator towards the steady state, computed for the density matrix and for a single quantum trajectory~\cite{plenio98a} for a system initially in the ground state.
The Wigner function of the oscillator shows an initial squeezing along two directions that is eventually confined in phase space due to the TLS nonlinearity (see Appendix~\ref{sec:apF}). Individual quantum trajectories reveal that the bimodal steady state consists of a cat state undergoing random phase flips due to single-phonon losses~\cite{minganti16a}, as shown in the last two columns of Fig.~\ref{fig:4}(b), that capture two times, before and after a single-phonon emission event. Each of these cat states has an extremely long lifetime, surviving with fidelities $F>0.99$ for times longer than a millisecond (see Appendix~\ref{sec:apG}).

\emph{Transient non-classical states.}--- The high quality factors of state-of-the-art nanoresonators~\cite{ghadimi18a} allows for non-classical states to develop and evolve in timescales of tenth of a second before every trace of coherence is washed out. We show this by plotting the evolution of the ``cattiness'' ${\mathcal C=\mathcal{N}(\rho)/\mathcal{N}(\rho_\mathrm{cat})}$, defined by dividing the integrated negative parts of the Wigner function of the state by that of a reference cat state~~\cite{everitt14a}, so that $\mathcal C>0$ only for nonclassical states and $=1$ for cat states. The results shown in Fig.~\ref{fig:5} demonstrate that we can observe unambiguous nonclassical features lasting up to seconds even with state-of-the-art setups~\cite{ghadimi18a}.
Several routes to detect these quantum states are discussed in Appendix~\ref{sec:apE}. Once in the steady state, a feedback protocol has been proposed~\cite{minganti16a} in order to enhance the decay rate only when the system is in one of the two possible cat states, and therefore stabilize the system in the other. We note that the combination of recently developed single-phonon detectors~\cite{cohen15a,riedinger16a} and the optical control of decay via active cooling makes the system proposed here an attractive platform to implement such feedback protocols, e.g., switching between two effective quality factors---by changing the driving amplitude of the cooling laser---whenever a single phonon is detected.

\begin{acknowledgments}
The authors kindly acknowledge P. B. Li, C. Navarrete-Benlloch, X. Hu and A. Miranowicz for useful discussions and comments. C.S.M acknowledges funding from a Short-Term Grant from the Japanese Society for the Promotion of Science (JSPS) and from the Marie Sklodowska-Curie Fellowship QUSON (Project  No. 752180).  J.P. was supported by the RIKEN Incentive Research Project Grant No. FY2016. 
F.N. is supported in part by the MURI Center for Dynamic Magneto-Optics via the Air Force Office of Scientific Research (AFOSR) (FA9550-14-1-0040), Army Research Office (ARO) (Grant No. 73315PH), Asian Office of Aerospace Research and Development (AOARD) (Grant No. FA2386-18-1-4045), Japan Science and Technology Agency (JST) (the ImPACT program and CREST Grant No. JPMJCR1676), Japan Society for the Promotion of Science (JSPS) (JSPS-RFBR Grant No. 17-52-50023), RIKEN-AIST Challenge Research Fund, and the John Templeton Foundation.\end{acknowledgments}

\bibliographystyle{mybibstyle}

\bibliography{Sci,books,arXiv}


\section*{Appendix}

\subsection{Magnetic field simulations}
\label{sec:apA}
For the case of $\mathrm{Dy}$, we have considered a saturation magnetization of $\mu_0 M_\mathrm{s}$\SI{=3.7}{\tesla}~\cite{scheunert14a}, exchange stiffness $A=$~\SI{1.5d-12}{\joule.\metre^{-1}}~~\cite{saraiva10a}, and a magnetic damping  $\alpha=$~\num{0.036}~\cite{woltersdorf09a}; for $\mathrm{Co}$, we take $\mu_0 M_\mathrm{s}$\SI{=1.79}{\tesla}~\cite{cantu15a}, $A=$~\SI{3.1d-11}{\joule.\metre^{-1}}~\cite{manfred_book03a} and $\alpha=$~\num{0.005}~\cite{schoen16a}; and for $\mathrm{FeCo}$, $\mu_0 M_\mathrm{s}$\SI{=2.4}{\tesla}~\cite{schoen16a}, $A=$~\SI{1.7d-11}{\joule.\metre^{-1}}~\cite{liu08a} and $\alpha=$~\num{e-4}~\cite{schoen16a}.
Note that, regarding the fabrication of these nanopillars, outstanding control on the size and shape can be achieved by using a variety of techniques, such as molecular beam epitaxy~\cite{pigeau12a} or focused electron-beam-induced deposition (FEBID)~\cite{gavagnin14a}. We calculated the magnetic field generated by the cylinders using MuMax3~\cite{vansteenkiste14a}, a finite differences, open-source solver of the Landau-Lifshitz-Gilbert equation~\cite{gilbert04a}. 

\begin{figure}[b!]
\begin{center}
\includegraphics[width=0.95\columnwidth]{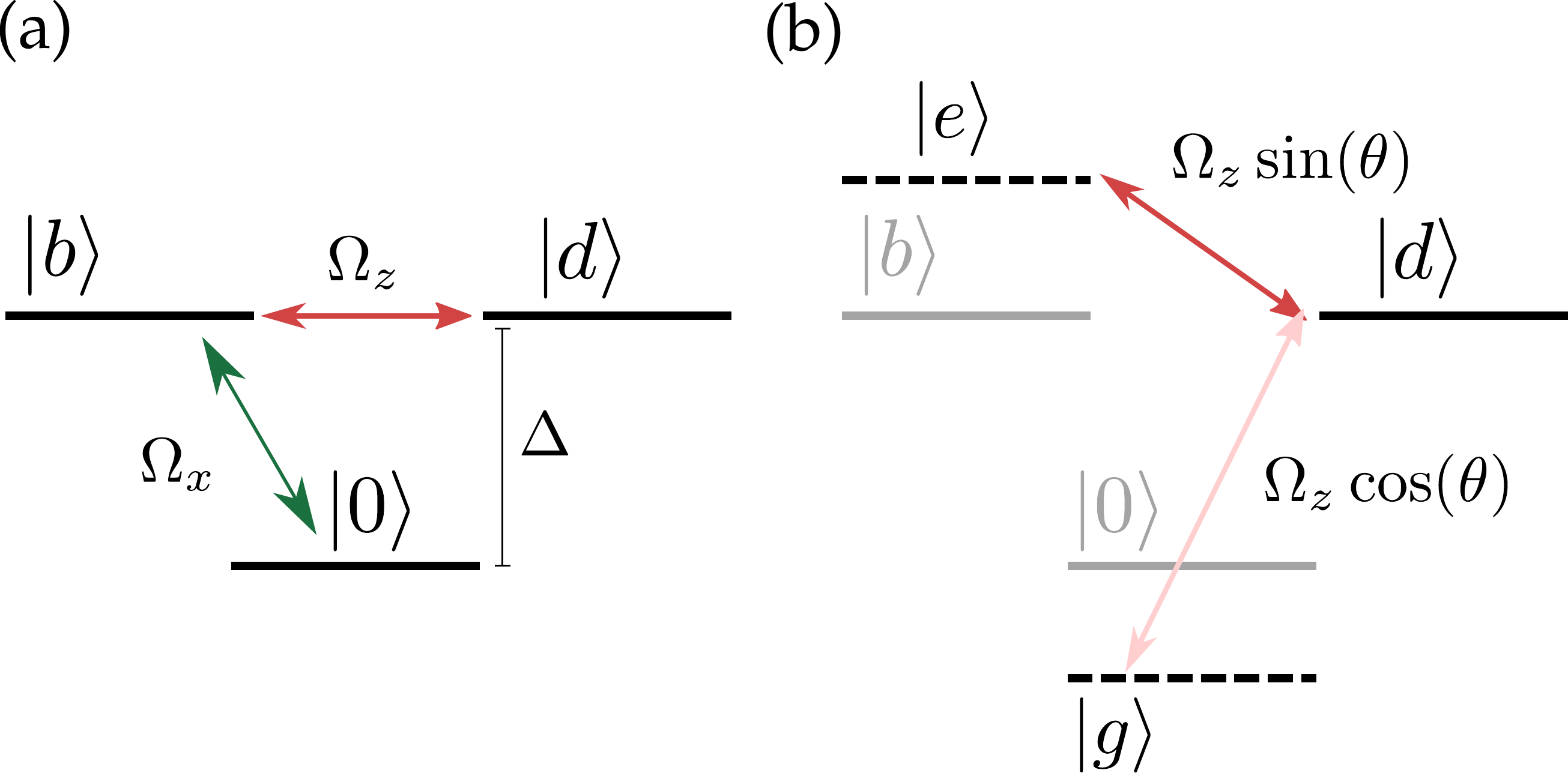}
\end{center}
\caption{Procedure to obtain an effective TLS. (a) Bare energy levels defined in the basis of bright and dark states and in the frame rotating with $\omega_x$, see Eq.~\eqref{eq:initial-two-phonon-H}. $|b\rangle$ is coupled to $|0\rangle$ through the microwave magnetic field in the $x$ axis, and to $|d\rangle$ through the magnetic field in the $z$ axis. (b) The microwave field dresses the states $|b\rangle$ and $|0\rangle$. In the basis of dressed energy levels, the states $|e\rangle$ and $|d\rangle$ define an effective TLS, driven by the field in $z$ axis. If $\Delta \gg \Omega_x$, $\sin(\theta)\approx 1$.  }
\label{fig:1-appendix}
\end{figure}

\subsection{Master equation and quantum trajectory simulations}
\label{sec:apB}
Simulation of the system dynamics in a dissipative environment has been performed following two different techniques: master equation simulations and the method of quantum trajectories. 

For the master equation simulations, we solved numerically the differential equations that govern the evolution of the density matrix:
\begin{equation}
\dot{\rho} = -i[H,\rho] + (\gamma_m n_\mathrm{th}/2)\mathcal{L}_a[\rho] + (\gamma_z/2)\mathcal{L}_{\sigma^\dagger\sigma}[\rho], 
\end{equation}
where $\mathcal{L}_O[\rho] \equiv 2 O\rho O^\dagger - O^\dagger O\rho - \rho O^\dagger O$. This was done by truncating the Hilbert space setting a maximum number $N$ of phonons in the oscillator. For the calculations done in this manuscript, $N=100$ was enough to guarantee convergence for the highest values of driving considered. 

The method of quantum trajectories yields a stochastic evolution of a pure wavefunction, which averaged over many different realizations provides the same predictions as the master equation for the density matrix. At every finite time step $dt$, for each element of the type $(\gamma_i/2) \mathcal L_{0_i}[\rho]$ in the master equation, the wavefunction $|\psi(t)\rangle$ can randomly undergo a quantum jump with probability $p_i = \gamma_i \langle\psi(t)|O_i|\psi(t)\rangle dt$ that transforms the system as
\begin{equation}
|\psi(t+dt)\rangle \propto O_i|\psi(t)\rangle
\end{equation}
(under proper normalization). The occurrence of a jump is determined by generating a random number $r_i\in[0,1]$ at each time step, so that the jump occurs whenever $r<p_i$ ($dt$ must be chosen small enough so that, at every time step, $p_i\ll 1$). When no jump occurs, the wavefunction evolves as
\begin{equation}
|\psi(t+dt)\rangle \propto e^{-i H_\mathrm{eff} dt}|\psi(t)\rangle,
\end{equation}
where $H_\mathrm{eff}$ is a non-Hermitian Hamiltonian, $H_\mathrm{eff}\equiv H-i\sum_i (\gamma_i/2)O_i^\dagger O_i$.

\subsection{Derivation of the two-phonon, driven Jaynes-Cummings Hamiltonian}
\label{sec:apC}
Here we define an effective TLS in a way very similar to the one outlined in Refs.~\cite{rabl09a,li16a}. Our starting point is the Hamiltonian:

\begin{multline}
H = D S_z^2+ \Omega_x \cos(\omega_x t) S_x + \Omega_z \cos(\omega_z t) S_z \\+ \omega_m a^\dagger a + g_2(a^\dagger + a)^2 S_z\,.
\label{eq:initial-two-phonon-H}
\end{multline}
 By working in the basis of bright and dark states
\begin{eqnarray}
\ket{b}=\frac{1}{\sqrt{2}}(\ket{-1}+\ket{1}),\\
\ket{d}=\frac{1}{\sqrt{2}}(\ket{-1}-\ket{1}),
\end{eqnarray}
and assuming, $\Omega_x \ll \omega_x \sim D$, we can perform a rotating wave approximation and apply a unitary transformation ${U = \exp[i\omega_x t (\op{b}{b}+\op{d}{d})]}$ to Eq.~\eqref{eq:initial-two-phonon-H} in order to move to a rotating frame where the time dependence with $\omega_x$ is eliminated:
\begin{multline}
H = \Delta (\op{b}{b} + \op{d}{d}) + \omega_m a^\dagger a + \left[\Omega_x\op{0}{b} \phantom{(a^\dagger + a)^2} \right.\\
\left. + \Omega_z \cos(\omega_z t)\op{b}{d} + g_2 (a^\dagger + a)^2 \op{b}{d}+ \mathrm{h.c.}\right] ,
\end{multline}
with $\Delta \equiv D-\omega_x $. The driving term with $\Omega_x$ can be removed by working in the dressed basis of states $\ket{g}$ and $\ket{e}$:
\begin{eqnarray}
\ket{g} = \sin(\theta )\ket{0} - \cos(\theta)\ket{b},\\
\ket{e} = \cos(\theta)\ket{0} + \sin(\theta)\ket{b},
\end{eqnarray}
with $\cos(\theta) = 1/\sqrt{1+\xi^{-2}}$, ${\sin(\theta) = 1/\sqrt{1+\xi^2}}$, and $\xi = \Omega_x/(\Delta/2 + R)$, where $R = \sqrt{\Omega_x^2 + (\Delta/2)^2}$.
In the limit $\Delta \approx 0$, we have $\omega_{gd} = \omega_{de} = \Omega_x$. This makes it evident that we should take the opposite limit $\Delta \gg \Omega_x$, since this will increase the difference between $\omega_{gd}$ and $\omega_{de}$, allowing us to spectrally isolate one of these transitions as an effective TLS. In that case, ${\omega_{gd} \approx \Delta+\Omega_x^2/\Delta}$, and $\omega_{de} \approx \Omega_x^2/\Delta$, so that $\omega_{gd}\gg \omega_{de}$, and $\sin(\theta) \approx 1$. By defining an effective TLS with lowering operator $\sigma \equiv \op{d}{e}$ and transition energy $\omega_\sigma = \omega_{de}$, and assuming $\Omega_z \ll \omega_\sigma$, we can make a rotating wave approximation to eliminate fast-rotating terms and perform a final unitary transformation $U = \exp[i \omega_z t (\sigma^\dagger \sigma +  a^\dagger a/2)] $ to remove the remaining time dependence, yielding the driven, two-phonon Jaynes-Cummings Hamiltonian of Eq.~(2) in the main text. Here and in the main text, we define $\Omega \equiv \Omega_z$ to lighten the notation. The scheme presented here is sketched in Fig.~\ref{fig:1-appendix}.  The derivation will be valid for ${\Omega_x \ll\{\Delta,\omega_x\}}$, ${\Omega \ll \Omega_x}$, and ${g_2 \ll \omega_m}$, with $\Delta \equiv D-\omega_x $. Since ${D = 2\pi\times}$~\SI{2.88}{\giga\hertz}, a sensible choice of parameters is $\omega_x =2\pi\times$~\SI{1}{\giga\hertz}, giving $\Delta = 2\pi\times$~\SI{1.88}{\giga\hertz}. Since we want $\omega_\sigma = 2\omega_m \approx 2\pi\times$~\SI{3.6}{\mega\hertz}, this implies  a value $\Omega_x \approx 2\pi\times$~\SI{82}{\mega\hertz}, which fulfills the conditions above and sets the maximum limit for the driving $\Omega$. 
\subsection{First order magnetic gradient effects due to imperfect alignment}
\label{sec:apD}
\begin{figure}[t!]
\begin{center}
\includegraphics[width=0.9\columnwidth]{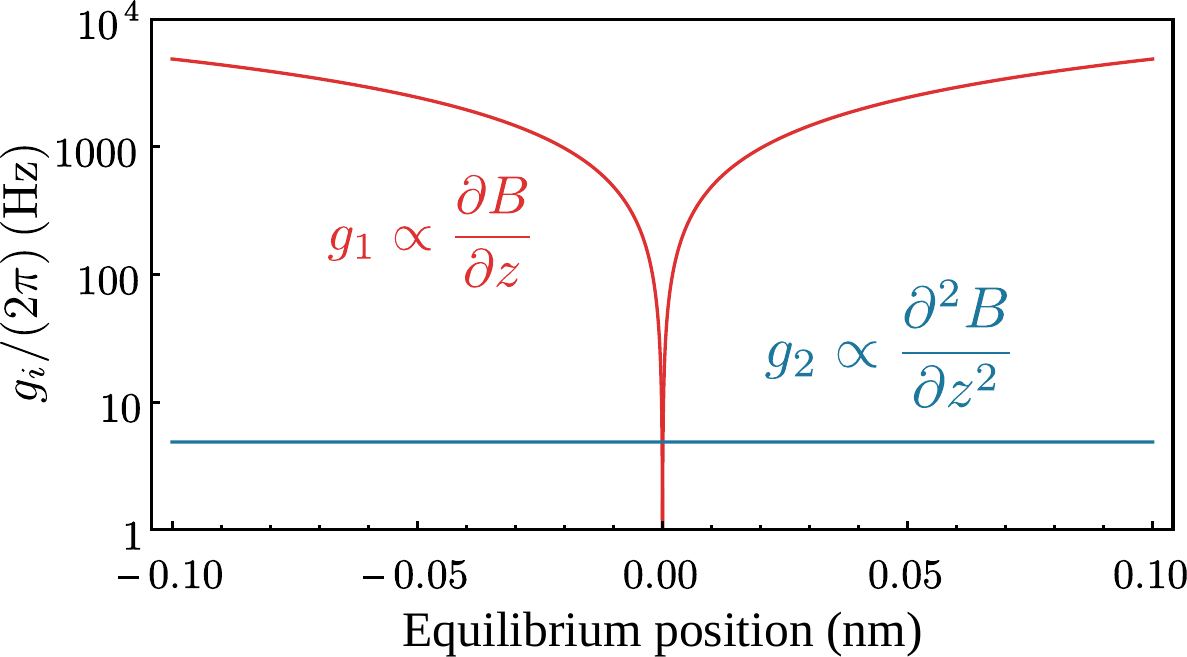}
\end{center}
\caption{First and second-order coupling rates as a function of the equilibrium position of the oscillator with respect to the center of the gap between the magnets.}
\label{fig:2-appendix}
\end{figure}

In the main text, we considered a geometry in which the equilibrium point of the oscillator lies exactly at the center of the gap between two nanomagnets, yielding a null first-order gradient of the magnetic field and therefore a pure quadratic coupling. However, it is important to address to which extent unavoidable deviations from a perfectly aligned situation might render noticeable effects due to the induced coupling through first-order gradients. In the simulation depicted in Fig.~\ref{fig:2-appendix}, we observe that a misalignment of \num{0.1}~nm is able to induce first-order couplings in the range of \si{\kilo\hertz}. Taking into account that we are considering mechanical modes with frequencies $\omega_m \sim $~\si{\mega\hertz} and fixing the two-phonon resonant condition $\omega_\sigma \approx 2\omega_m$, we see that first-order gradient terms of the kind $g_1 (a+a^\dagger)(\sigma+\sigma^\dagger)$ will rotate as $\sim \exp[{\pm i \omega_m t}]$ and can therefore be neglected beside unimportant frequency shifts (e.g., even for $g_1$ as high as \SI{100}{\kilo\hertz}, one can still achieve full two-phonon Rabi oscillations governed by $g_2/(2\pi)=$\SI{5}{\hertz} by tuning the TLS frequency to $\omega_\sigma = \omega_m-\lambda$, with $\lambda \approx $\SI{3.54}{\kilo\hertz}). 

By tuning the TLS in resonance with the mechanical mode, $\omega_\sigma = \omega_m$, the first-order coupling can be used as well to stabilize the resonator close to the middle point. The variation in $g_1$ as the oscillator is moved would yield different responses of the TLS, which could be used in a feedback loop to correct the position of the resonator. It has been demonstrated that picometer stability can be achieved by adding feedback control to piezo-actuators via spectroscopy arrangements, which can be readily obtained by monitoring the NV center light emission~\cite{walder15a,hummer16a}.\\[1pt]

\subsection{Detection of mechanical non-classical states}
\label{sec:apE}

In the main text, we have focused on the generation of non-classical states of motion. In an experimental implementation, it is vital do to have a scheme to detect and reconstruct such states. There is a great body of work regarding the reconstruction of mechanical states~\cite{vanner15a}; here, we comment on the route consisting of a displacement and a phonon number measurement. One possible way to see that this technique allows to reconstruct the quantum state is to note that the Wigner function can be written as 
\begin{equation}
W(\alpha,\alpha^*)=2\pi\Tr{\hat D^\dagger(\alpha)\hat \rho \hat D(\alpha)\hat P},
\end{equation}
with $\hat D$ the displacement operator and $\hat P$ the parity operator. Through phonon number measurements, we can obtain the expected value of the parity operator for different displaced states and reconstruct the Wigner function.

\begin{figure*}[t!]
\begin{center}
\includegraphics[width=1\textwidth]{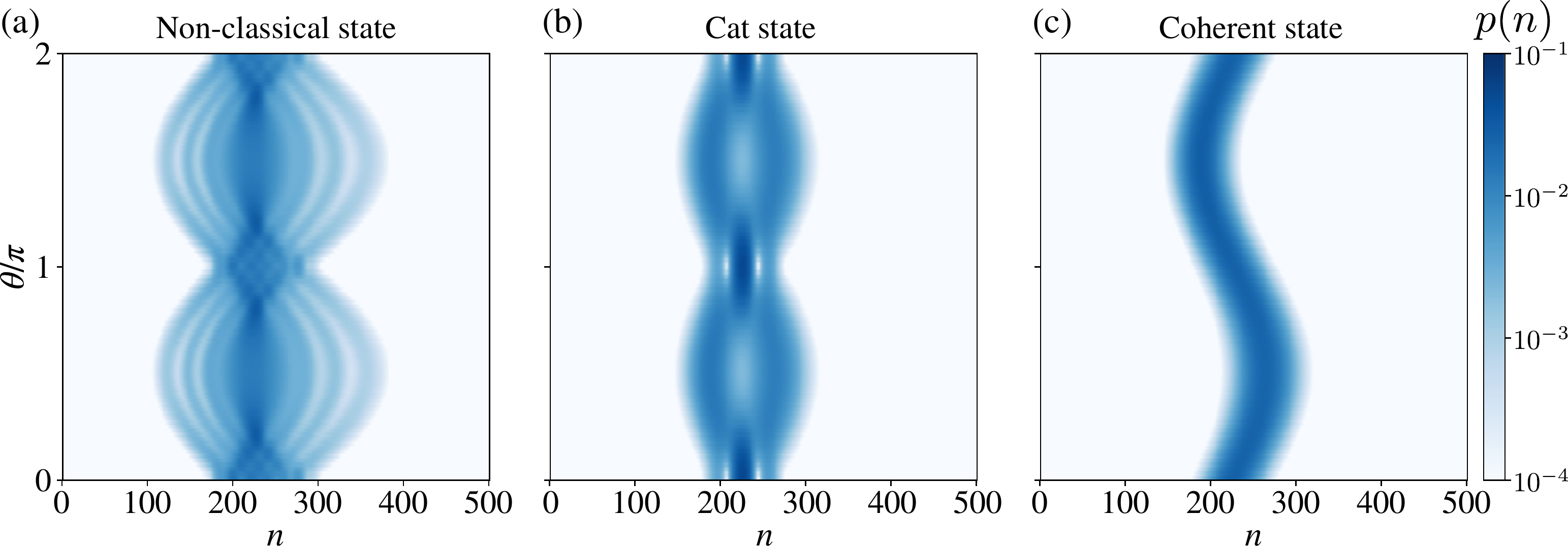}
\end{center}
\caption{Simulation of a state-reconstruction measurement.  The phonon number distribution $p(n)$ is measured mechanical state displaced by a fixed amplitude $|\alpha|=15$ for a different set of displacement angles $\theta$. The states used here are: (a) a non-classical mechanical state generated by our system (the same as in the inset of Fig.~5 in the main text); (b) a cat state with the same phonon population as in (a); (c) a coherent state with the same phonon population as in (a). This data can be used to retrieve the quantum state of the system via reconstruction algorithms.}
\label{fig:3-appendix}
\end{figure*}

We can also picture the number measurement on a displaced state as an homodyne measurement; by considering the displaced annihilation operator $\tilde a = a+\alpha$, with $\alpha=|\alpha|e^{i\theta}$, the resulting number operator is 
\begin{equation}
\tilde a^\dagger \tilde a = a^\dagger a + a^\dagger \alpha + \alpha^* a + |\alpha|^2.
\end{equation}
For $|\alpha|^2\gg \langle a^\dagger a \rangle$, we obtain that the number measurement of the displaced state minus an offset $|\alpha|^2$ measures the quadrature amplitude $Q_\theta$:
\begin{equation}
(\tilde a^\dagger \tilde a -|\alpha|^2)/(\sqrt{2}|\alpha|) \approx (a e^{-i\theta}+a^\dagger e^{i\theta})/\sqrt{2} =Q_\theta.
\end{equation}
Measured over one complete cycle in $\theta$, the quadrature amplitudes $Q_\theta$ provide tomographically complete information about the quantum system~\cite{lvovsky09a,vogel89a}. Figure~\ref{fig:3-appendix} shows a simulation of the proposed measurement; the phonon number distribution of a mechanical state displaced by a fixed amplitude $\alpha$ is measured for a set displacement angles $\theta$. The difference between the resulting data for the distinct states considered is apparent even to the naked eye; this information can be used to infer the quantum state through multiple reconstruction algorithms, like maximum likelihood or entropy maximization~\cite{vanner15a}. 

The qubit-resonator coupling can also be used in order to perform state tomography of the mechanical oscillator. First order effects allow us to go from a resonant two-phonon coupling regime to a resonant or dispersive one-phonon coupling regime by tuning the TLS energy out of the two-phonon resonance. This would allow, for instance, to measure the state of the oscillator generated by the two-phonon interaction by suddenly switching the TLS energy to a regime of dispersive interaction governed by $g_1$, which can be used to employ techniques of state reconstruction via displacement and number measurement through Ramsey interferometry of the qubit~\cite{haroche_book06a,deleglise08a,vanner15a}. 

\subsection{Confined dynamics in phase space}
\label{sec:apF}
The role of nonlinearity brought by the TLS is to limit the dynamics to a region of the phase space of the oscillator. This is clearer if we represent Eq.~(2) in the basis that diagonalizes the driven-TLS Hamiltonian~\cite{sanchezmunoz18a}, considering, for simplicity, the resonant case $\omega_\sigma = \omega_z$:
\begin{equation}
H = \Omega \tilde\sigma_z + (\omega_m-2\omega_z) a^\dagger a + \frac{g_2}{2}[{a^\dagger}^2(\tilde\sigma - \tilde\sigma^\dagger + \tilde\sigma_z) +\mathrm{h.c.}]
\end{equation}
At high driving, $\Omega \gg g_2$, the terms proportional to ${a^\dagger}^2(\tilde \sigma - \tilde\sigma^\dagger)+\mathrm{h.c.}$ are counter-rotating and do not contribute to the dynamics provided that $n_a g_2 \ll \Omega$, where $n_a$ is the phonon population in the oscillator. In this case, the evolution under the terms ${a^\dagger}^2\tilde \sigma_z + \mathrm{h.c.}$ is decoupled for the two eigenstates of $\tilde\sigma_z$, $|\pm\rangle$, and takes the form of a squeezing operation along the angle $\pm \pi/4$. Due to this squeezing operation, the phonon population of the oscillator grows. Once the population reaches a value such that $n_a g_2 \approx \Omega$, the counter-rotating terms enter into action and distort the evolution. They act as a barrier in phase space, preventing a small initial population from growing past a given threshold, in close similarity the physics of confined quantum Zeno dynamics~\cite{raimond10b,raimond12a,signoles14a}. 
This is shown in Fig.~4 in the main text, where we depict the evolution, from the moment the driving is turned on, of a system initially in its ground state. In order to gain insight in the driven-dissipative nature of the dynamics, we show the phonon population and Wigner functions of the reduced cavity system, computed both from the density matrix and from the wavefunction of a single quantum trajectory~\cite{plenio98a}. Initially, the TLS is in its ground state, which is described in the dressed basis as a linear superposition $|g\rangle \propto |+\rangle + |-\rangle$. Therefore, the mechanical mode evolves in a superposition of being squeezed along the $\pi/4$ and $-\pi/4$ axes, yielding a cross-like pattern in phase space, as shown in the first column of Fig.~4(b). After that, the counter-rotating terms enter into action and the squeezing is distorted, yielding a ribbon-like pattern in a confined region of phase space. Finally, the interplay between the coherent evolution and dissipation in the resonator yields a steady Wigner function with two lobes associated to the coherent states $|\pm i \sqrt{\Omega/g_2}\rangle$.

\subsection{Long-lived cats in a quantum trajectory}
\label{sec:apG}
\begin{figure}[b!]
\begin{center}
\includegraphics[width=0.95\columnwidth]{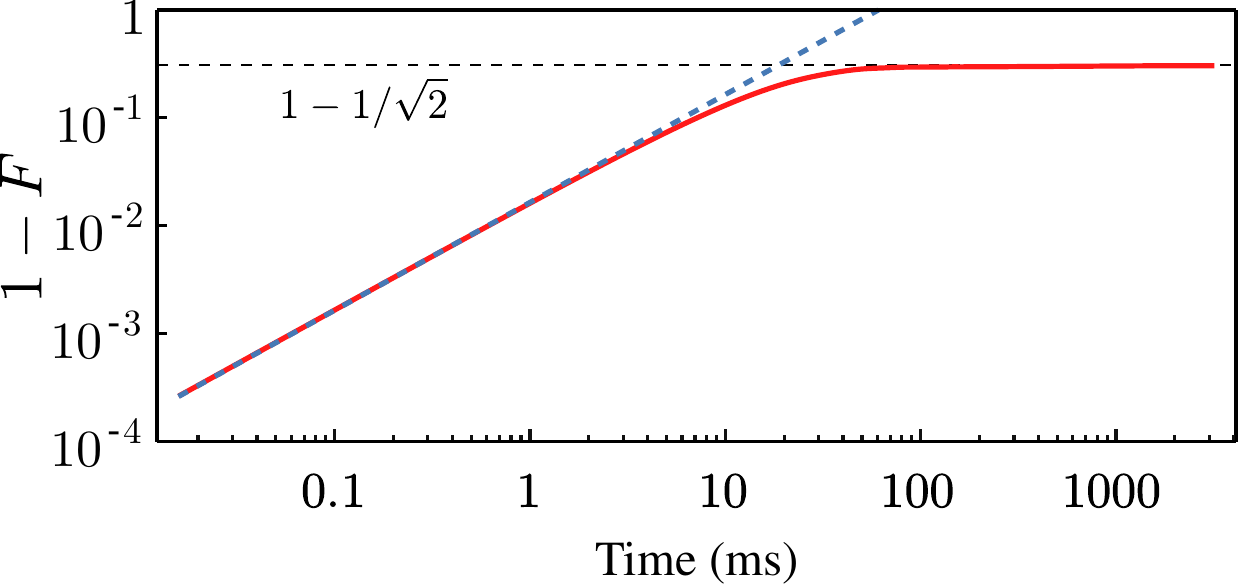}
\end{center}
\caption{Fidelity $F$ to a cat state versus time for an initial state chosen from a random time ($t=0$) in the quantum trajectory, once the steady-state limit is reached. The initial state resembles a cat with fidelity $F>0.999$. The blue, dashed line corresponds to the expression for $F \approx 1-\frac{1}{2}\gamma_m n_\mathrm{th} n_a t$, which is valid for short times. For long times, the fidelity tends to $F =1/\sqrt{2}$. Simulation parameter, those of Figs.~3-4.}
\label{fig:fidelity}
\end{figure}

When considering a single quantum trajectory, any cat state in which the system is found to be remains stable for very long times, even in the absence of any feedback protocol. The reason is that they are only affected by the random quantum jumps that flip their phase. At any time, the probability to undergo a jump during a small time interval $dt$ is $p = \gamma_m n_\mathrm{th} n_a dt$. Therefore, if the system is initialized in one of the two cat states that compose the mixed steady state, this state remains stable with a fidelity:
\begin{equation}
F(t) \approx 1-\frac{1}{2}\gamma_m n_\mathrm{th} n_a t,
\end{equation}
where we considered time intervals shorter than the phonon lifetime, i.e., $t < 1/(\gamma_m n_\mathrm{th}n_a)$. Since phonon lifetimes can reach hundreds of seconds in oscillators with high quality factors~\cite{ghadimi18a}, a cat state in this system can in fact be extremely long lived. This is shown in Fig.~\ref{fig:fidelity}, where we selected a pure state of the quantum trajectory at a random time (once the evolution is stationary) that is very close to a cat state, let it evolve as a mixed state under the master equation, and computed the fidelity to a cat state with the same population as the initial state. This shows that the cat state can be maintained stable in this system with a fidelity $F>0.99$ for times $\sim 1$~ms,  during which it can be used as a resource for quantum applications~\cite{ralph03,lund08,joo11,facon16a,albert16a}.

\end{document}